\documentclass[twoside]{article}
\usepackage[latin1]{inputenc}
\usepackage{t1enc}
\usepackage{a4}
\usepackage{tabularx,subfigure}
\usepackage{graphicx,epsf}

\makeatletter
\newcommand\figcaption{\def\@captype{figure}\caption}
\newcommand\tabcaption{\def\@captype{table}\caption}
\makeatother

\def\bref{\vspace{4pt}\noindent\hangindent=10mm}

\begin{document}

\setcounter{figure}{0}
\setcounter{section}{0}
\setcounter{equation}{0}
\setcounter{footnote}{1}

\begin{center}
{\Large\bf
Modelling the spectral energy distribution of galaxies from the ultraviolet to 
submillimeter}\\[0.7cm]

Cristina C. Popescu$^{1,}$\footnote{Otto Hahn Fellow of the Max Planck 
Institut f\"ur Astronomie, K\"onigstuhl 17, 69117 Heidelberg, 
Germany}$^{,}$\footnote{Research Associate, The Astronomical Institute of 
the Romanian Academy, Str. Cu\c titul de Argint 5, Bucharest, Romania} 
and Richard J. Tuffs$^{4}$\\[0.17cm]
$^1$ The Observatories of the Carnegie Institution of Washington, \\
813 Santa Barbara Str., Pasadena, CA 91101, USA, \\
popescu@ociw.edu\\
$^4$ Max Planck Institut f\"ur Kernphysik, Saupfercheckweg 1,\\
 69117 Heidelberg, Germany\\
\end{center}

\vspace{0.5cm}

\begin{abstract}
\noindent{\it
We present results from a new modelling technique which can account for
the observed optical/NIR - FIR/submm spectral energy distributions (SEDs)
of normal star-forming galaxies in terms of a minimum number of essential
parameters specifying the star-formation history and geometrical
distribution of stars and dust. The model utilises resolved optical/NIR
images to constrain the old stellar population and associated dust, and
geometry-sensitive colour information in the FIR/submm to constrain the
spatial distributions of young stars and associated dust.
It is successfully applied to the edge-on spirals NGC~891 and
NGC~5907. In both cases the young stellar population powers the bulk
of the FIR/sub-mm emission. The model also accounts for the observed
surface brightness distribution and large-scale radial brightness profiles 
in NGC~891 as determined using the Infrared Space Observatory (ISO) at 
170 \& 200$\,{\mu}m$ and at 850$\,{\mu}m$ using SCUBA.
}
\end{abstract}

\section{Introduction}

Historically, almost all our information about the 
current and past star-formation properties of galaxies has been based 
upon spatially integrated measurements in the
ultraviolet (UV), visible and near-infrared (NIR) spectral regimes. However, 
star-forming galaxies 
contain dust which absorbs some fraction of the emitted starlight, 
re-radiating it predominantly in the far-infrared (FIR)/sub-millimeter(submm) 
range. The true significance of this process even for ``normal'' 
(i.e. non-starburst) galaxies has been revealed by observations of a 
representative sample of late-type Virgo cluster galaxies with the 
ISOPHOT instrument on board the Infrared Space Observatory (ISO). 
These showed the dust emission to typically account 
for 50 percent of the bolometric output of these systems, with a spectral 
peak generally lying between 100 and 250$\,{\mu}m$ (Tuffs et al. 2002;
Popescu et al. 2002).

In view of this, the measurement of current and past 
star-formation in galaxies - and indeed of the universe as a whole -
requires a quantitative understanding of the 
role different stellar populations play in powering the FIR/submm
emission. For this both optical and FIR/submm data needs to be used, as they
contain complementary  information about the distribution of stars and dust. 

On the one hand, optical data probes the colour and spatial 
distribution (after correction for extinction) of the photospheric
emission along sufficiently transparent lines of sight. This is
particularly useful to investigate older, redder stellar populations
in galaxian disks with scale heights larger than that of the dust. 
On the other hand, grains act as test particles probing the strength 
and colour of ultraviolet (UV)/optical interstellar radiation fields. 
This constitutes
an entirely complementary constraint to studies of photospheric emission.
In the FIR, grains are moreover detectable over almost the full range of optical
depths present in a galaxy. At least part of this regime is  inaccessible
to direct probes of starlight, especially at shorter wavelengths, even 
for face-on systems. This particularly applies to light from young stars
located in, or close by, the dust clouds from which they formed, since 
a certain fraction of the light is locally absorbed. Furthermore, 
there is at least a possibility that most of the remaining UV and
even blue light from young stars that can escape into the disk might
be absorbed by diffuse dust there. A combined analysis of the whole 
UV-optical\-/FIR\-/submm spectral energy distribution (SED) of galaxies 
seems to be a promising way to constrain the problem.

Here we present the results from a new  tool  for modelling  the  
optical-FIR/submm SED  of galaxies (Popescu et al. 2000b, 
Misiriotis et al. 2001). 
Our model predicts the SED as  a function of intrinsic SFR, star 
formation history   (from   population    synthesis   calculations)   
and   dust content. This tool includes solving the radiative-transfer 
problem for
a  realistic  distribution  of  absorbers  and  emitters,  considering
realistic  models  for  dust  (taking  into  account  the  grain-size
distribution  and   stochastic  heating   of  small  grains)  and  the
contribution of HII regions.   The model addresses the fundamental and
controversial question  of optical thickness  in the disk  of galaxies
and  can account  for the  detailed UV-optical-FIR-submm  SED  with an
absolute  minimum  of independent  physical  variables, each  strongly
constrained by data.

\section{An overview of SED modelling in the literature}

The SED models of galaxies have been built using the tools and results from 
previous studies of the key ingredients required by such models: radiative 
transfer codes and dust emission models. Radiative transfer codes were 
initially developed to account for the UV-optical/NIR appearance of 
galaxies, their extinction properties, and the role the spatial distribution 
of stars and dust play in shaping the SED. These codes
used either analytical methods (Kylafis \& Bahcall (1987) or Monte Carlo 
simulations (Witt \& Gordon 1996, 2000, Ferrara et al. 1999, 
Gordon et al. 2001). In parallel, 
dust models have been built to model extinction curves and the heating 
and emission of grains in different environments. These include the
studies of Mathis et al. (1977), Draine \& Lee (1984), 
Draine \& Anderson (1985), Dwek (1986), Guhathakurta \& Draine (1989), 
D\'esert et al. (1990), Laor \& Draine (1993), Weingartner \& Draine (2001).

Applications of such methods to study the SED of galaxies have been 
initiated, as expected, in the optical/NIR regime. Thus, in their fundamental 
work, Kylafis \& Bahcall (1987) applied a radiation transfer modelling 
technique to 
edge-on systems, where the scale height of the stars and dust extinction
can be directly constrained. The technique was subsequently applied on several 
edge-on galaxies by Xilouris et al. (1997, 1998, 1999).  Radiative
transfer codes in combination with observations of nearby edge-on galaxies were
also used by Ohta \& Kodaira (1995), Kuchinski et al. (1998) and Matthews \&
Wood (2001). 

A further step in the SED modelling was to include the FIR-submm spectral
regime in the overall analysis. Different tools have been proposed starting 
with the pioneering works of Xu \& Buat (1995) and Xu \& Helou (1996). They used
radiative transfer codes for an assumed ``sandwich'' configuration of dust and 
stars and  considered in detail the relative contribution of the
non-ionising UV photons and the optical photons in heating the grains.
However, these calculations did not incorporate a model for the dust grain 
emission, nor for radial variations in the absorbed radiation in the disk, and
therefore could not account for the shape of the FIR SED.  

Recently, there have been several works modelling the SED of galaxies from the 
 UV to submm, which considered more realistic geometries and dust models, and which
have started to make use of the new observational data becoming available at
longer FIR/submm wavelengths. In parallel with these self-consistent models
a series of empirical or semi-empirical approaches for modelling the
SED from UV to submm wavelengths have been adopted, mainly for statistical 
applications (Devriendt et al. 1999, Dale et al. 2001). 

Self consistent calculations of the SED over the entire
spectral range have been developed slowly, due to the large amount of
computational time required both for the radiative transfer and for the 
calculation of the probability distribution of temperatures of dust grains, 
especially of the small grains, which undergo large temperature fluctuations. 
To this should be added that it
is only now becoming possible to obtain resolved images covering the MIR to
FIR and submm spectral range, even for nearby galaxies. Such observations 
are needed for constraining SED models. 
Recent self-consistent SED modelling was presented by  Silva et
al. (1998). They applied photometric and chemical evolution models of 
galaxies to explain the SED of both normal and starburst galaxies. Their model
considers radiative transfer in both the molecular clouds and in the diffuse
ISM, and includes a consistent treatment of dust emission and stochastic 
heating of small grains. It can thus reproduce reasonably well the SED of the 
studied galaxies, in both the optical and FIR spectral range and is therefore 
suited to describe the general shape of the volume-integrated SED. The model
has however more free parameters than the generally available observational 
constraints and the solution obtained may not be unique. The model is also
less adequate in describing the intrinsic distributions of stars 
and dust because of its simplified geometry. For example the
model does not consider different scale heights and scale lengths for
the stellar and dust distribution. Furthermore, all stellar populations are
constrained to have
the same scale heights, including the young stellar population which is
known to reside in a very thin stellar disk. The radiative transfer is only an
approximation, and is rigorously applicable only to an infinite homogeneous
medium, and there is no treatment of anisotropic scattering. 

Bianchi et al. (2000) attempted to model NGC~6946 from the UV to FIR
using a 3D Monte Carlo radiative-transfer code for a simplified
geometry of emitters (a single stellar disk). They concluded that the
total FIR output is consistent with an optically thick solution. 
This result should be considered with caution as their model makes a number of
simplifications. For example they did not consider a size distribution for 
grains and there is no consistent treatment of grain heating and emission. 
Accordingly, there is no
calculation of stochastic heating for small grains, but rather a MIR correction
calculated according to the model of D\'esert et al. (1990). Furthermore the
model did not consider the contribution of localised sources within 
star-forming complexes. This resulted in a poor fit of the FIR SED and a 
failure to reproduce the IRAS flux densities.  

Several studies have been dedicated to modelling the SED of starburst 
galaxies. 
Recent work on modelling the UV to submm emission was presented by 
Efstathiou et al. (2000) for starburst galaxies, which were treated as
an ensemble of optically thick giant molecular clouds (GMCs) centrally
illuminated by recently formed stars. The model has a proper treatment of dust
emission, including stochastic heating of small grains, and it
allows for the star-forming complexes to evolve with time, based on a physical
model of the dusty HII region phase and of the supernova phase. The 
star-forming complexes are modelled assuming spherical symmetry, which, 
as remarked
by the authors, may not be true in the later stages of the GMC evolution.  
Also there is no further transfer of radiation between GMCs, meaning that the
stellar light is not allowed to escape the GMC. The FIR spectrum is fitted for an exponential decaying star formation
rate, where the age and time constant of the star-burst (or in some cases of
two star-bursts) are fitted parameters. This modelling technique
successfully reproduced the observed SED of M~82 and NGC~6090, though for the
latter case there is an ambiguity between the existence of an older burst
and a diffuse component not considered in the model. Obviously this method is
suitable for starburst dominated galaxies and can bring interesting insights 
into the ages of the stellar populations, but cannot be applied to normal 
``quiescent'' disk galaxies dominated by emission from the diffuse 
interstellar radiation field.

Bekki et al. (2000)  studied the time evolution of the SED in starburst 
galaxies based on numerical simulations that can analyse simultaneously
dynamical and chemical evolution, structural and kinematical properties. This
method is rather unique and can be used to study mergers, for example, or other
physical processes related to galaxy formation. The price to pay for this
complexity is the quite rudimentary treatment of the radiative 
transfer and dust emission: scattering has been neglected 
from the radiative transfer, there is no grain size distribution and 
no treatment of stochastic heating of small grains. Therefore, their model can 
be used to indicate trends in the SED evolution rather than to give 
quantitative predictions and is suitable for numerical simulations.  

As a general remark, most of the work described above has been concentrated
in fitting the volume-integrated emission of the galaxies at different spectral
wavelengths and has not been constrained by morphological information from 
images. A more robust knowledge of the intrinsic distributions of
stellar populations and dust, and ultimately of the SF history in
galaxies can be gained if also the surface brightness distribution
of optical and even FIR emission can be analysed. The method we present here 
constrains the problem by using the whole information in the brightness 
distribution. This approach has been only used in the optical/NIR regime, by
Xilouris et al. (1997, 1998, 1999). With
the ISO and SCUBA data we are now able to make predictions also 
for the appearance of the FIR/submm images and compare the predicted maps with 
the observed ones. In the future SIRTF observations will extend
knowledge of the FIR morphology in galaxies and will bring new constraints for 
the SED modelling methods.

\section{Our model}

Star-forming galaxies are fundamentally inhomogeneous,
containing highly obscured massive star-formation regions, as well as more 
extended structures harbouring older stellar populations which may 
be transparent or have intermediate optical depths to starlight.
Accordingly, our model divides the stellar population into an 
``old'' component (considered to dominate the output in B-band 
and longer wavelengths) and a ``young'' component 
(considered to dominate the output in the non-ionising UV).

\subsection{The ``old'' stellar population}

The ``old'' stellar population is generally observed to have
scale heights of several hundred pc in rotationally
supported spiral galaxies, a result which can be physically 
attributed to the increase of the kinetic temperature of stellar 
populations on timescales of order Gyr due to encounters 
with molecular clouds and/or spiral density waves. This means that the ``old'' 
stellar population can be constrained from resolved optical and near-IR 
images via the modelling procedure of Xilouris et al. (1999). The procedure 
uses the technique for
solving the radiation transfer equation for direct and multiply scattered
light for arbitrary geometries by Kylafis \& Bachall (1987). 

For edge-on systems these calculations completely determine the scale heights 
and lengths of exponential disk representations of the old stars 
(the ``old stellar disk'') and associated diffuse dust 
(the ``old dust disk''), as well as a 
dustless stellar bulge. This process is feasible for edge-on
systems since the scale height of the dust is less than that of the 
stars. For face-on systems the scale height cannot be fixed by the optical data
alone, and must be determined through consideration of its effect on the FIR
SED. 

The extinction law of the dust can be directly determined from the
observed optical/near-IR wavelength, since the calculation is done
independently at each wavelength. In all cases so far,
an extinction law consistent with that predicted from the graphite/silicate
mix of Laor \& Draine (1993) and the $a^{-3.5}$ grain size distribution of 
Mathis Rumpl \& Nordsieck (1977) has been found (Xilouris et al. 1999).

\subsection{The ``young'' stellar population}

The ``young'' stellar population is also specified by an exponential
disk, which we shall refer to as the ``young stellar disk''. 
Invisible in edge-on systems, 
its scale height is constrained to be $90\,$pc (the value for the Milky Way)
and its scale length is equated to that of the ``old stellar disk'' in B-band.
The emissivity of the ``young stellar disk'' is
parameterised in terms of the current star formation rate ($SFR$)
by relating the non-ionising UV emission to $SFR$ using the
population synthesis models of Bruzual \& Charlot (2001) for $Z=Z_{\odot}$, 
a Salpeter initial mass function, a mass cut-off of $100\,{\rm M}_{\odot}$,
and an exponential decrease of the $SFR$ with time, with a time constant  
$\tau\,=\,5\,Gyr$.

A second exponential dust disk of grain mass 
$M_{\rm dust}$ - the ``second dust disk'' is associated with the 
young stellar population. This is needed to account for the 
observed submm emission from edge-on disk galaxies, 
which cannot be reproduced by models containing only the old dust 
disk determined from the optical images
(Popescu et al. 2000b, Misioritis et al. 2001). It is
constrained to have the same scale length and height as that of the 
young stellar disk. This second dust disk is assumed to be composed of
grains with the same graphite/silicate and $a^{-3.5}$ grain size 
distribution as for the old dust disk. All the exponential disks for 
young and old stars and dust in the model are truncated at three times 
the exponential scale length. Because two disks of dust are required 
for the model, we refer to it as to the ``two-dust-disk'' model.

The current star-formation rate ($SFR$) and mass of the second
dust disk ($M_{\rm dust}$) are the first two primary 
free parameters of the model to determine the FIR/submm radiation.
They both relate to the smooth distribution of stars and dust
in the second disk. A third primary parameter, $F$, is included to
account for inhomogeneities in the distributions of dust and stars 
in the young stellar disk. $F$ is defined as the fraction of non-ionising UV
which is locally absorbed in HII regions around the massive stars.
It determines the additional likelyhood of
absorbtion of non-ionising UV photons due to autocorrelations 
between an inhomogeneous distribution of young stars and 
parent molecular clouds. Astrophysically, this arises because at any
particular epoch some fraction of the massive stars have not had time 
to escape the vicinity of their parent molecular clouds. Thus, $F$
is related to the ratio between the distance a star travels in its lifetime
due to it's random velocity and the typical dimensions of star-formation 
complexes.

\subsection{Calculation of dust emission}

The first step in the calculation of the dust emission is to
determine the radiation energy density of the unabsorbed
light versus wavelength from the non-ionising UV to the near-IR
for the diffuse radiation field. This is done for trial values 
of the mass of dust in the second dust disk $M_D$ by solving the 
radiation transfer equation for the observed optical wavelengths and 
three wavelengths in the non-ionising UV. (For edge on systems,
$M_D$ is independent of the dust mass in the old dust disk, which
is fixed by the optical observations). The actual energy density distribution
can then be determined for any trial combination of $SFR$ and $F$
without a further radiation transfer calculation. Both the
colour and intensity of the radiation field vary with position
in different ways according to the combination of $SFR$, $M_D$ and $F$.

The FIR/submm emission for each combination of $SFR$, $M_D$ and $F$ is 
then calculated for graphite and silicate
grains of size $a$ immersed in the radiation field at each point of
a grid of positions in the galaxy, including an explicit 
treatment of stochastic emission. 
Subsequently we integrate over the entire galaxy to obtain the FIR-submm 
SED of the diffuse disk emission.
Prior to comparison with observed FIR-submm SEDs, an empirically
determined spectral template for the HII regions, scaled according 
to the value of $F$, must be added to this calculated spectral 
distribution of diffuse FIR emission. 

Due to the precise constraints on the distribution of stellar emissivity in 
the optical-NIR and the distribution and opacity
of dust in the ``old dust disk'' yielded by the radiation 
transfer analysis of the highly resolved optical-NIR images, coupled with the
simple assumptions for the distribution of the young stellar 
population and associated dust, our model has just three free 
parameters - $SFR$, $F$ and $M_{\rm dust}$. These fully determine
the FIR-submm SED, and allow a meaningful comparison with broad-band
observational data in the FIR/submm, where, in particular for distant
objects, typically only a few spectral sample points for the
spatially integrated emission are available. 
The parameters are strongly coupled, but in general terms, 
$M_{\rm dust}$ is principally constrained by the 
submm emission, $SFR\,\times\,(1-F)$ by the bolometric FIR-submm output 
and the factor $F$ (in the absence of high resolution images and/or for edge-on
systems) by the FIR colour.

\section{Application to edge-on spiral galaxies}

Modelling edge-on spiral galaxies has several advantages, mainly when 
investigating them in the optical band. One advantage is that, in this view of 
a galaxy, one can easily
separate the three main components of the galaxy (i.e., the stellar disk, the
dust and the bulge). Another advantage is that the dust is very prominently 
seen in the dust lane, and thus its scalelength and scaleheight can be better 
constrained. A third advantage is that many details of a galaxy that are 
evident when the galaxy is seen face-on (e.g., spiral arms), are smeared out 
to a large degree when the galaxy is seen edge-on (Misiriotis et al. 2000). 
Thus, a simple model with relatively
few parameters can be used for the distribution of stars and dust in the
galaxy. However, in edge-on galaxies it is very difficult to see
localised sources (i.e., HII regions), in which the radiation can be locally
absorbed and thus not contribute to the diffuse radiation field. Also, if the
galaxy has a thin (young) stellar/dust disk, highly obscured by the dust lane 
in the plane of the galaxy, then this disk cannot be inferred from 
observations in the optical/NIR spectral range. As described in the previous
section, our model makes use of the information in the FIR-submm regime to
constrain this problem.

We first applied the above method to the well-known edge-on spiral 
galaxy NGC~891. This is one of the most extensively observed edge-on galaxies 
in the nearby universe, which makes it ideal for a verification of our 
modelling technique. We have also extended our SED modelling 
technique to four additional edge-on 
systems - NGC~5907, NGC~4013, UGC~1082 and 
UGC~2048 - with the aim of examining whether the features of the 
solution we obtained 
for NGC~891 might be more generally applicable. In this section we mainly 
show and discuss the results for NGC~891, and only 
briefly illustrate the solution for NGC~5907. 

The ``two-dust-disk'' model can successfully fit the
shape of the SED for both NGC~891 and NGC~5907. 
The best solution for NGC~891 (Fig.~1) has a central face-on V-band optical 
thickness 
$\tau^f_V=3.1$ and a corresponding non-ionising UV  luminosity 
$\sim 8.2\times 10^{36}$\,W. 
The luminosity of the diffuse component is $4.07\times 10^{36}$\,W, which 
accounts for $69\%$ of the observed FIR luminosity, and the luminosity of the 
HII component is $1.82\times 10^{36}$\,W, making up the remaining $31\%$ 
of the FIR luminosity. The best solution for  NGC~5907 (Fig.~2) has a central 
face-on optical depth in the optical band ${\tau}_{\rm v}=1.4$.
The total FIR-submm re-radiated 
luminosity of NGC~5907, obtained by integrating the ``two-dust-disk'' 
model SED, is $50.5\times 10^{35}$\,W out of which $27.0\times 10^{35}$\,W
is attributed to heating from the young stellar population.
Thus, about $40\%$ of the dust emission is powered by the old stellar 
population. The major difference between
NGC~891 and NGC~5907, on the basis of the ``two-dust-disk'' model, 
is that the spectrum of the former apparently allows for the existence of a 
larger contribution from HII regions (see Misiriotis et al. 2001 for a 
detailed discussion) - $F$ takes values of 0.22 and 0.10, respectively.
Such small values of $F$ are expected for ``normal'' galaxies,
in contrast to starburst systems, where the FIR/submm SEDs peak shortwards 
of $100\,\mu$m, and one would anticipate that $F$ would be closer to 
unity.

In both cases most of the
luminosity comes from the diffuse component, and the main heating source is
provided by the young stellar population. 
The relative contribution of optical and UV photons in 
heating the dust has been a longstanding question in the literature. 
Since we have a detailed calculation 

\hspace{-0.6cm}
\begin{minipage}[t]{\textwidth}
\includegraphics[scale=0.6]{sed.epsi}
\figcaption{The predicted SED of NGC~891 from the ``two-dust-disk'' model with 
$SFR=3.8\,{\rm M}_{\odot}$/yr, $F=0.22$ and 
$M_{\rm dust}=7\times 10^{7}\,{\rm M}_{\odot}$ in the second disk of dust. 
LH panel: the intrinsic emitted 
stellar radiation (as would have been observed in the absence of dust). 
RH panel: the re-radiated dust emission, with
diffuse and HII components plotted as 
dashed and dotted lines, respectively. The data  
(integrated over $\pm 225^{\prime\prime}$ in longitude), are 
from Alton et. al 1998 (at 60, 100, 450 \& 850\,$\mu$m), 
Gu\'elin et al. 1993 (at $1300\,\mu$m) and from Popescu et al. (2001) (at
170 \& 200\,$\mu$m).}

\vspace*{0.3cm}

\includegraphics[scale=0.6]{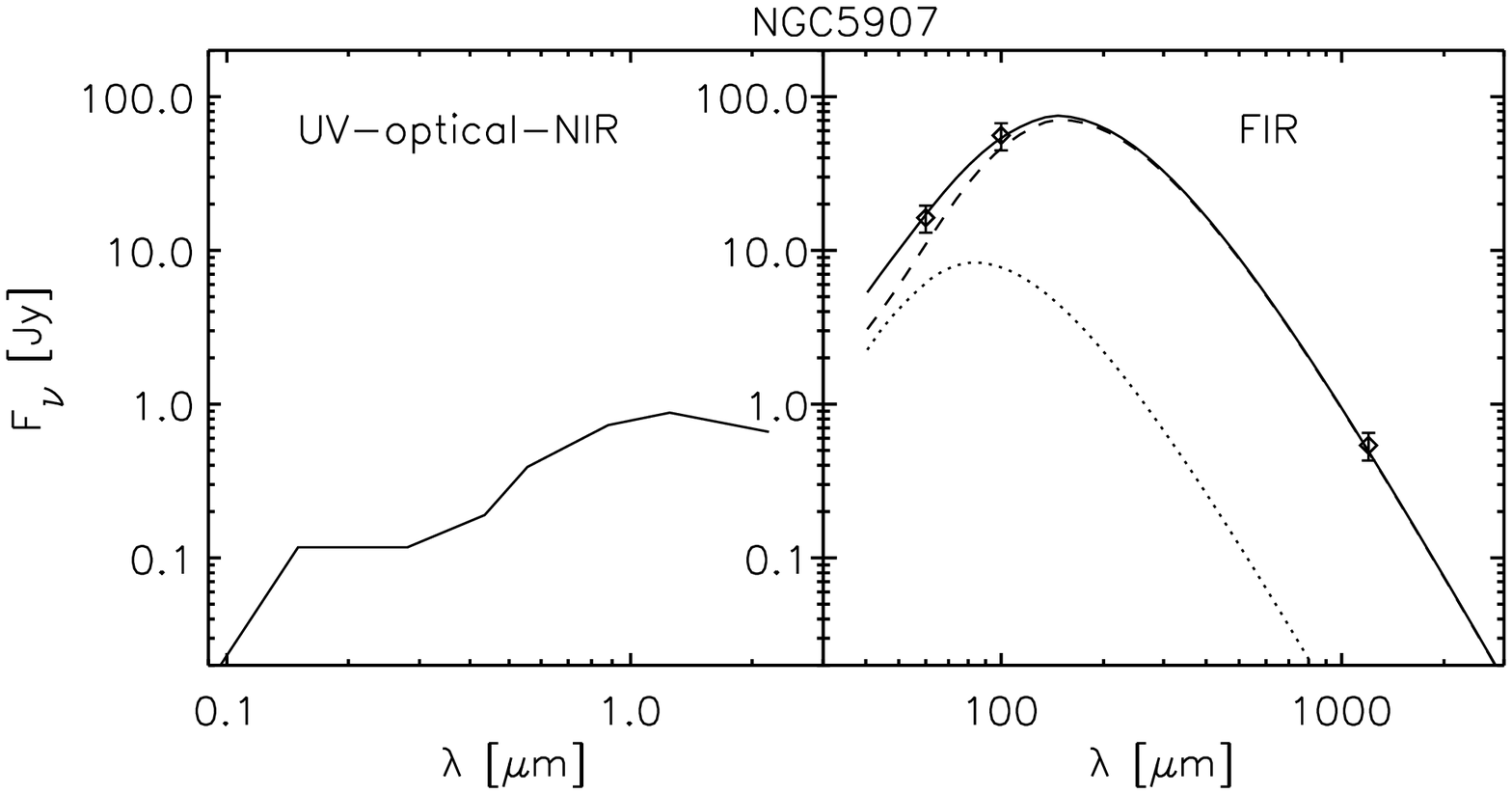}
\figcaption{The predicted SED of NGC~5907 from the ``two-dust-disk'' model 
with $SFR=2.2\,{\rm M}_{\odot}$/yr, $F=0.10$ and 
$M_{\rm dust}=4.5 \times 10^7\,{\rm M}_{\odot}$ in the second disk of dust. 
The legend is as in Fig.~1. The data are
from Young et al. (1989) (at 60 \&  100\,${\mu}$m), and from Dumke et al. 
(1997) (at 1200\,${\mu}$m)}
\end{minipage}
\\[\intextsep]

\hspace{-0.6cm}
\begin{minipage}[t]{\textwidth}
\includegraphics[scale=0.5]{spectrum1.epsi}
\figcaption
{The absolute contribution 
of the three stellar components to the FIR emission versus wavelength 
for the ``two-dust-disk'' model. Dashed-line: diffuse optical radiation 
($4000-22000$\,\AA); dashed-dotted line: diffuse UV 
radiation ($912-4000$\,\AA); dotted-line: HII regions.
The total predicted FIR SED is given by the solid line.
The data points are as for Fig.~1.}

\vspace*{0.3cm}

\includegraphics[scale=0.5]{fraction.epsi}
\figcaption
{The fractional contribution 
of the three stellar components to the FIR emission versus wavelength 
for the ``two-dust-disk'' model. The legend is as for Fig.~3 and the data 
points are as for Fig.~1.}
\end{minipage} 
\\[\intextsep]
of the absorbed energy over the whole spectral range and at each position in 
the galaxy, we can directly calculate which part of the emitted FIR 
luminosity from each volume element of the galaxy is due to the optical and 
NIR photons, and which part is due to the UV photons. In this way we can also 
predict the contribution of different stellar populations in heating the dust
as a function of FIR wavelength. Volume-integrated IR 
spectral components  

\hspace{-0.6cm}
\begin{minipage}[t]{\textwidth}
\includegraphics[scale=0.4]{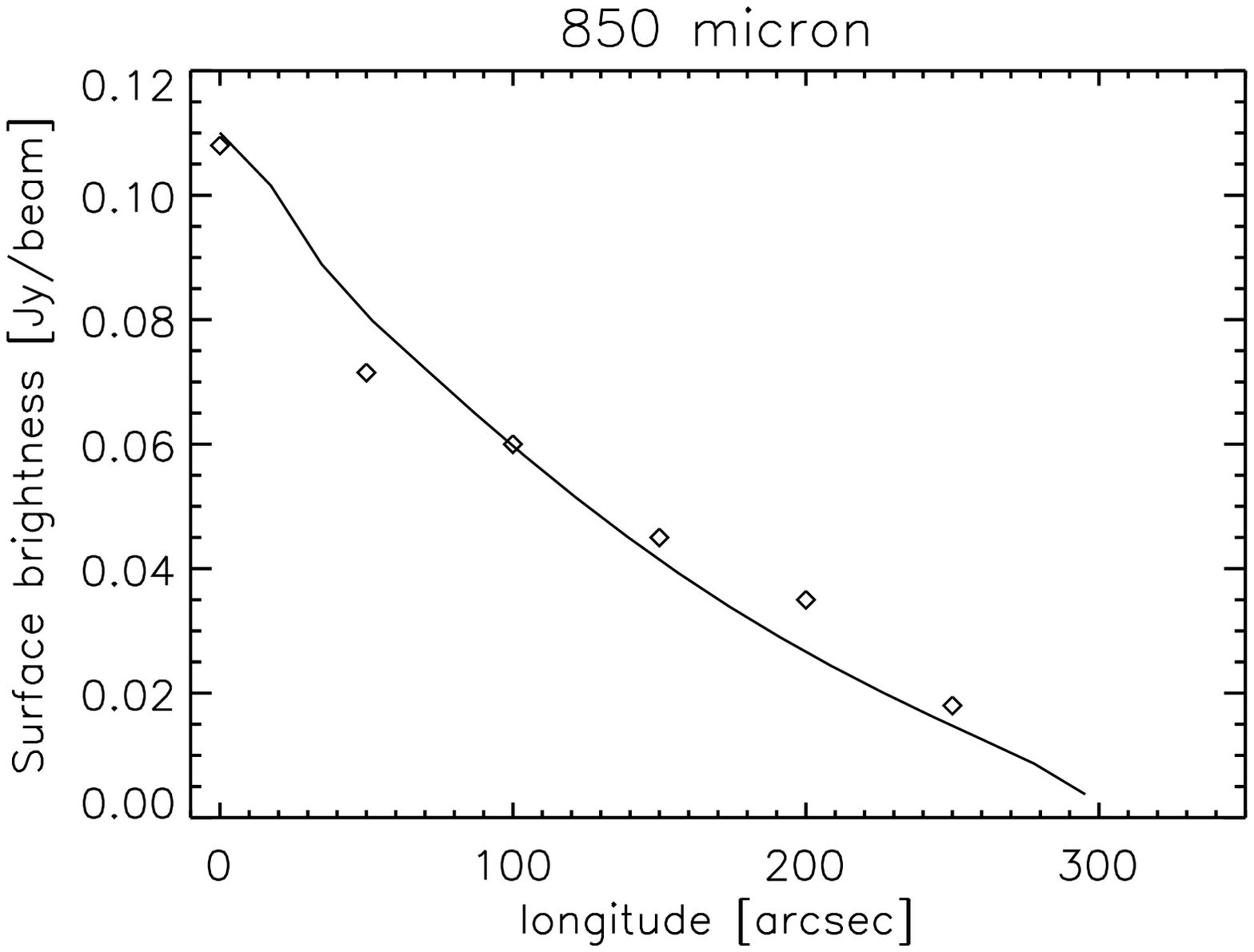}
\figcaption{
The averaged radial profile of NGC~891 at $850\,\mu$m 
from the diffuse component of the ``two-dust-disk'' model, plotted with the 
solid line. The profile is averaged over a bin width of $36^{\prime\prime}$, 
for a sampling of 
$3^{\prime\prime}$ and for a beam width of $16^{\prime\prime}$, in the same 
way as the observed averaged radial profile from Alton et al. (2000) 
(plotted with diamonds).}
\end{minipage} 
\\[\intextsep]
arising from re-radiated optical and UV light are presented
in Figs.~3,4 for the case of NGC~891. We note that the diffuse optical 
radiation 
field makes only a relatively small contribution to the total emitted dust 
luminosity. This is in
qualitative agreement with various statistical 
inferences linking FIR emission with young stellar populations, in particular 
the FIR-radio correlation. Our analysis predicts the predominance of 
UV-powered grain emission even in the submm range, which, in turn, would
predict a tighter FIR-radio correlation when the FIR luminosity integrated over
 the FIR-submm range is considered. This prediction has been recently
confirmed by Popescu et al. (2002), using the new ISOPHOT observations of a
complete sample of late-type Virgo cluster galaxies (Tuffs et al. 2002).

A more stringent test of the model is to compare its predictions for the
morphology of the dust emission with spatially resolved maps.
Because observed radial profiles were derived by Alton et al. 
(2000) using the SCUBA observations at $850\,\mu$m, we first attempt to
calculate the radial profiles at this wavelength and compare it with 
the observations (Fig.~5), which are mainly sensitive to 
dust column density.
We have found that in the
case of the ``two-dust-disk'' model there is a very good agreement between 
the model predictions and the observations, where the observed profiles were
mirrored for compatibility with the symmetry in our model. The predicted 
radial profile can be traced out to 300 arcsec radius (15 kpc), as also 
detected by the SCUBA.

Recent deep observations of NGC~891 with the ISOPHOT
instrument at 170 and 200\,${\mu}$m (Popescu et al. 2001) offer a 
still stronger test of the model, as the FIR emission here depends
on the distribution of both stellar luminosity and dust.

\hspace{-0.6cm}
\begin{figure}
\subfigure[]{
\includegraphics[scale=0.5]{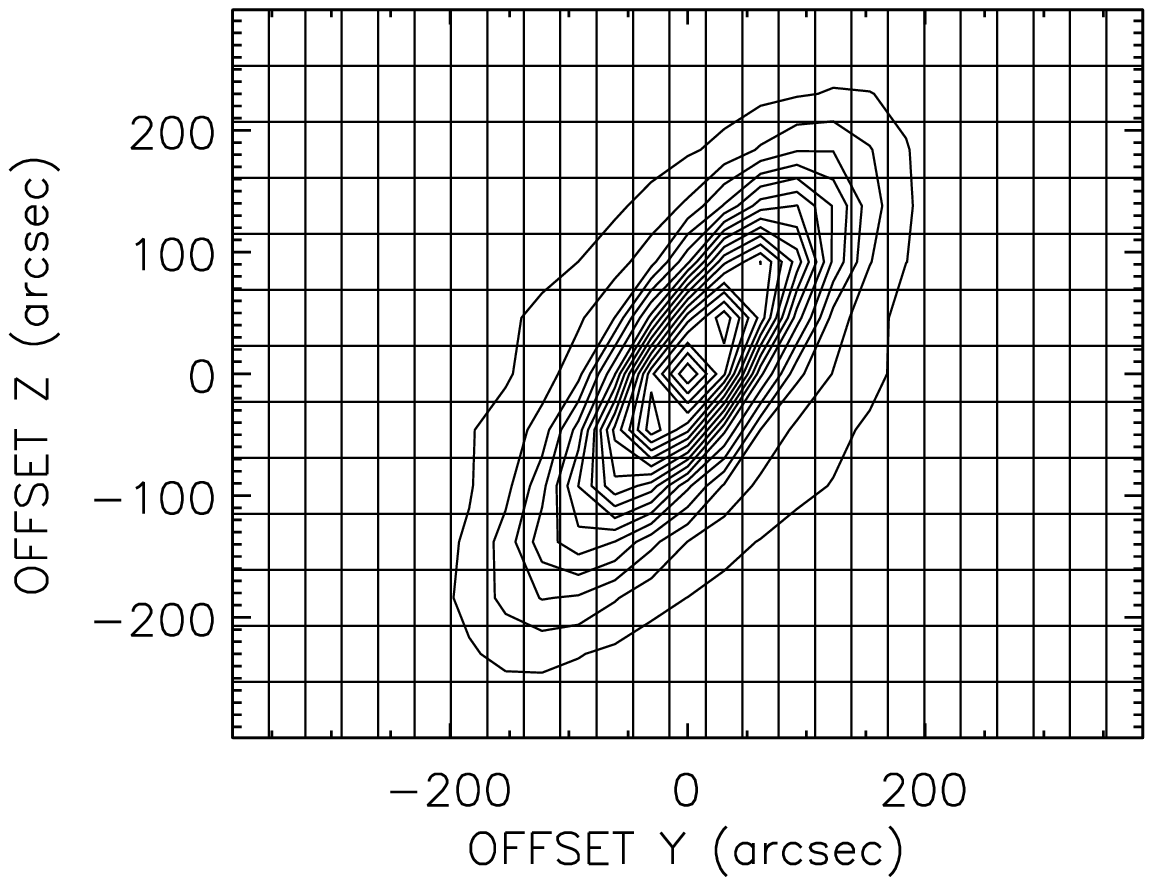}}
\subfigure[]{
\includegraphics[scale=0.5]{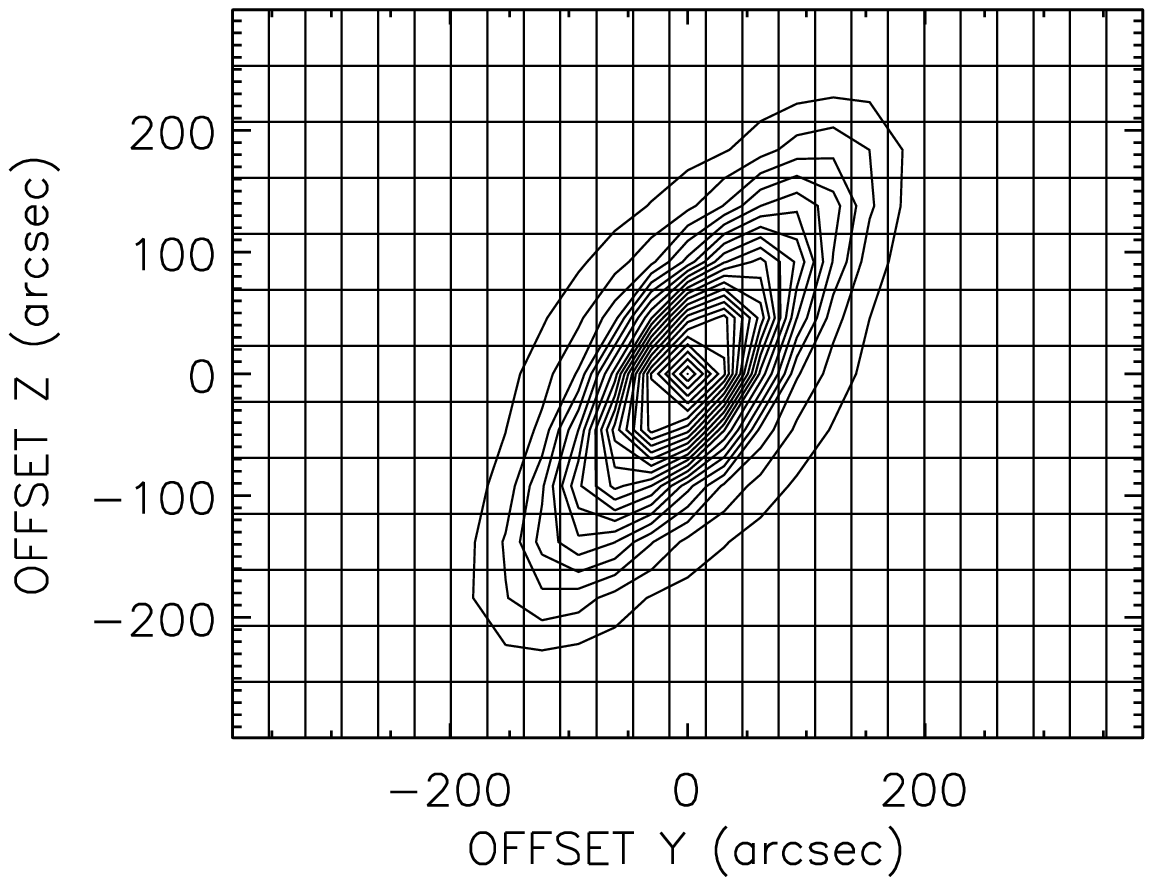}}
\subfigure[]{
\includegraphics[scale=0.5]{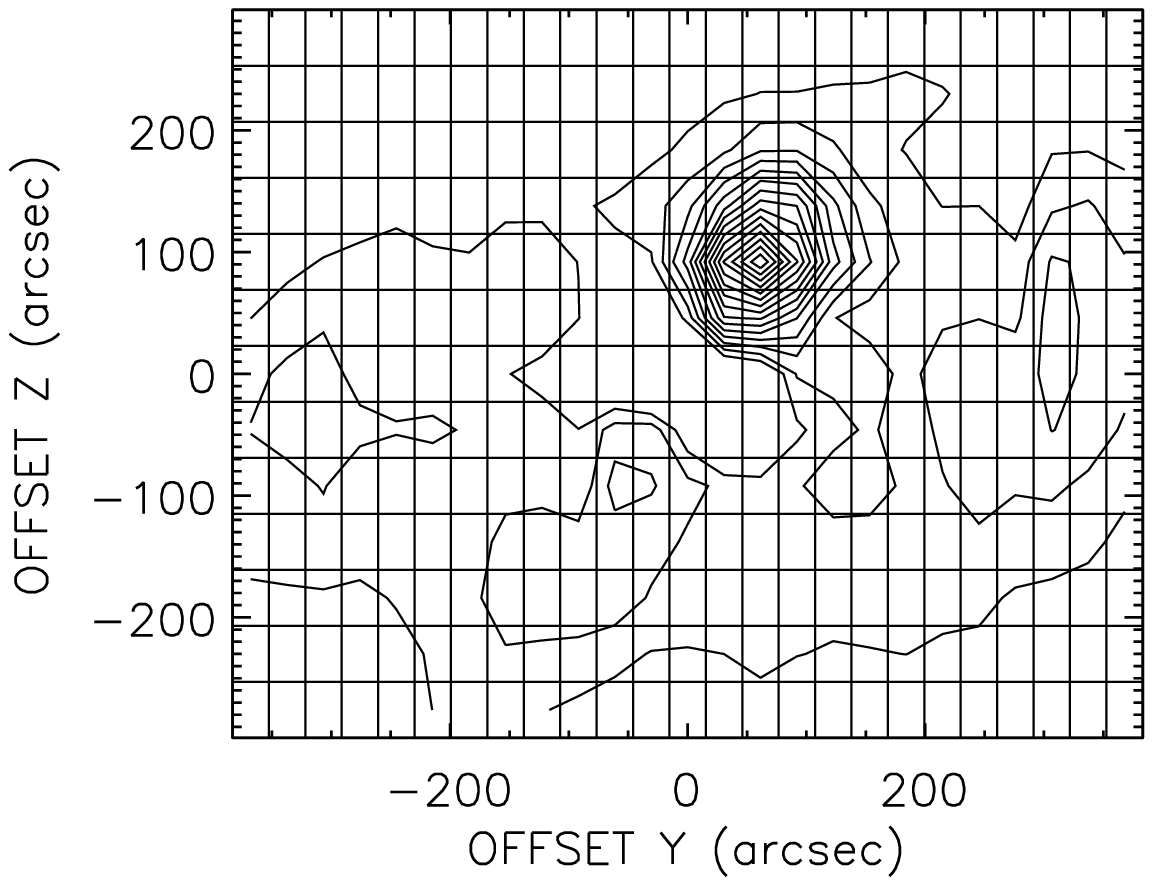}}
\hspace*{0.05cm}
\subfigure[]{
\includegraphics[scale=0.5]{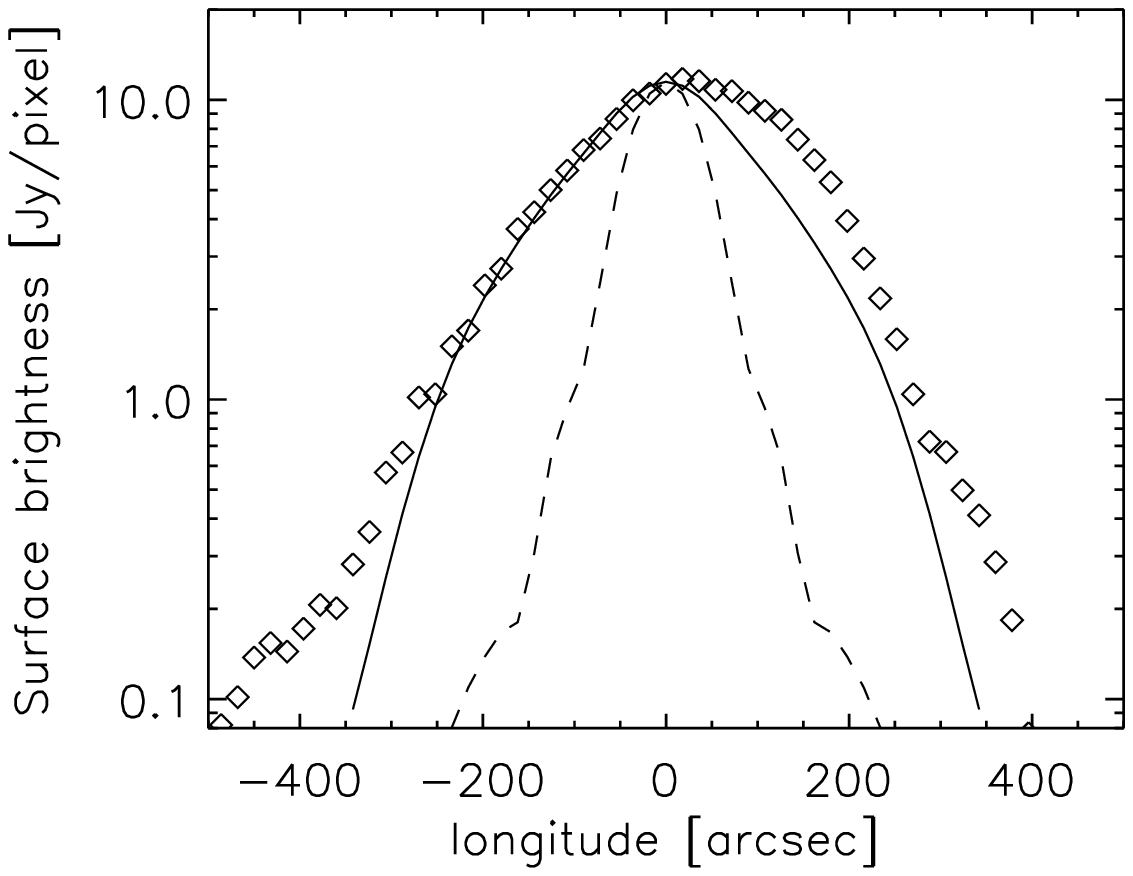}}
\caption{ 
a) Contour plot of the observed brightness distribution at
170\,${\mu}$m. The contours are plotted from 10.7 to 230.0\,MJy/sr in steps 
of 12.2\,MJy/sr. The grid indicates the actual measured sky positions 
sampled at 31 x 46 arcsec in the spacecraft coordinates Y and Z. 
b) Contour plot of the simulated diffuse brightness distribution at
170\,${\mu}$m. The contours are plotted from 10.4 to 239.7\,MJy/sr in steps 
of 10.4\,MJy/sr.
c) Contour plot of the observed minus simulated 
diffuse brightness distribution at 170\,${\mu}$m. The contours are plotted from 
2.2 to 46.1\,MJy/sr in steps of 3.1\,MJy/sr. The unresolved source from the
Northern side of the galaxy and the faint extended source from the Southern side
 account for $8\%$ of the total integrated flux density, in
agreement with our model prediction for the FIR localised sources at this
wavelength. 
d) The longitudinal profile (diamonds) observed by ISO
at $170\,\mu$m (Popescu et al. 2001), with a common bin width and sampling
interval of $18.4^{\prime\prime}$. North is towards positive longitude.
The solid line is the prediction from the diffuse component of the 
``two-dust-disk'' model and the dashed line the projected beam profile 
(FWHM 1.8 arcmin.) }
\end{figure}

Simulated FIR maps
were obtained using the actual pointing data to scan the diffuse disk 
model. The model map was then convolved with empirical PSFs derived from 
point source measurements. The comparison between the observed map (Fig.~6a) 
and the simulated one (Fig.~6b) show a remarkable agreement. To search for small
differences between the model and the observations, not detectable in the maps
due to the high dynamical range of the displayed data, we present in Fig.~6c the
residual map of the difference between the observed and the simulated map. The
main feature in the residuals is a localised, unresolved source in the Northern
side of the disk, with a peak of 46.1 MJy/sr. This localised source is probably  
a giant molecular cloud complex - associated with one of the spiral arms,
and not considered in the simulated map, which only includes the diffuse
component of the model. At this FIR wavelength our model predicts a $11\%$
contribution from the star-forming complexes. The integration of the unresolved
source gives a flux density of 9.4\,Jy, which is $6\%$ of the total integrated
flux density. Furthermore a faint extended source  is seen in the southern
side of the galaxy, of 4.0\,Jy integrated flux density. This makes another 
$2\%$ of the total integrated emission. Thus the faint localised sources seen in
the residual maps are in reasonable agreement with the prediction of our 
model, which reassures us that the template used in our model and scaled 
according to our model parameters, SFR and the F factor, are indeed a good 
representation for the galaxy. Apart from the two sources, a faint extended 
halo (at $\sim1\%$ brightness level) is seen extending at large heights 
perpendicular to the disk. The interpretation of this halo component is beyond 
the scope of this paper. 

A final comparison can be made between the projected radial profiles of the
simulated and observed maps. Due to the larger longitudinal coverage of the 
ISO data, which embraces the outer, asymmetrical HI disk 
(Swaters et al. 1997), this time we did not mirror 
the observed profile. Again, a very good agreement between the
model prediction and the observed profiles can be seen in Fig.~6d
for the Southern side of the galaxy. The localised FIR source
from the northern side of the residual map becomes prominent in the
radial profiles as well. There also 
seems to be an excess of FIR emission at radii larger than 
300$\,^{\prime\prime}$, not reproduced by our model. 
We interpret this result as indicative of a dust disk larger 
than considered by our model, in which all dust disks are truncated 
at three scale lengths of the B-band stellar disk. A finer grid of 
models with varying truncation of the scale length 
may be needed to reproduce the faint FIR emission at 
large galactocentric radii.

\section{The SFRs derived from SED modelling}

To statistically evaluate our results for SFR, we compare the SFR
characteristics of the 5 edge-on galaxies modelled by us with the larger 
sample of 61 galaxies with inclinations less than 75 degrees which were 
studied by Kennicutt (1998) on the basis of H$_{\alpha}$ measurements.

\begin {figure}[!t]                                                    
\includegraphics[scale=0.4]{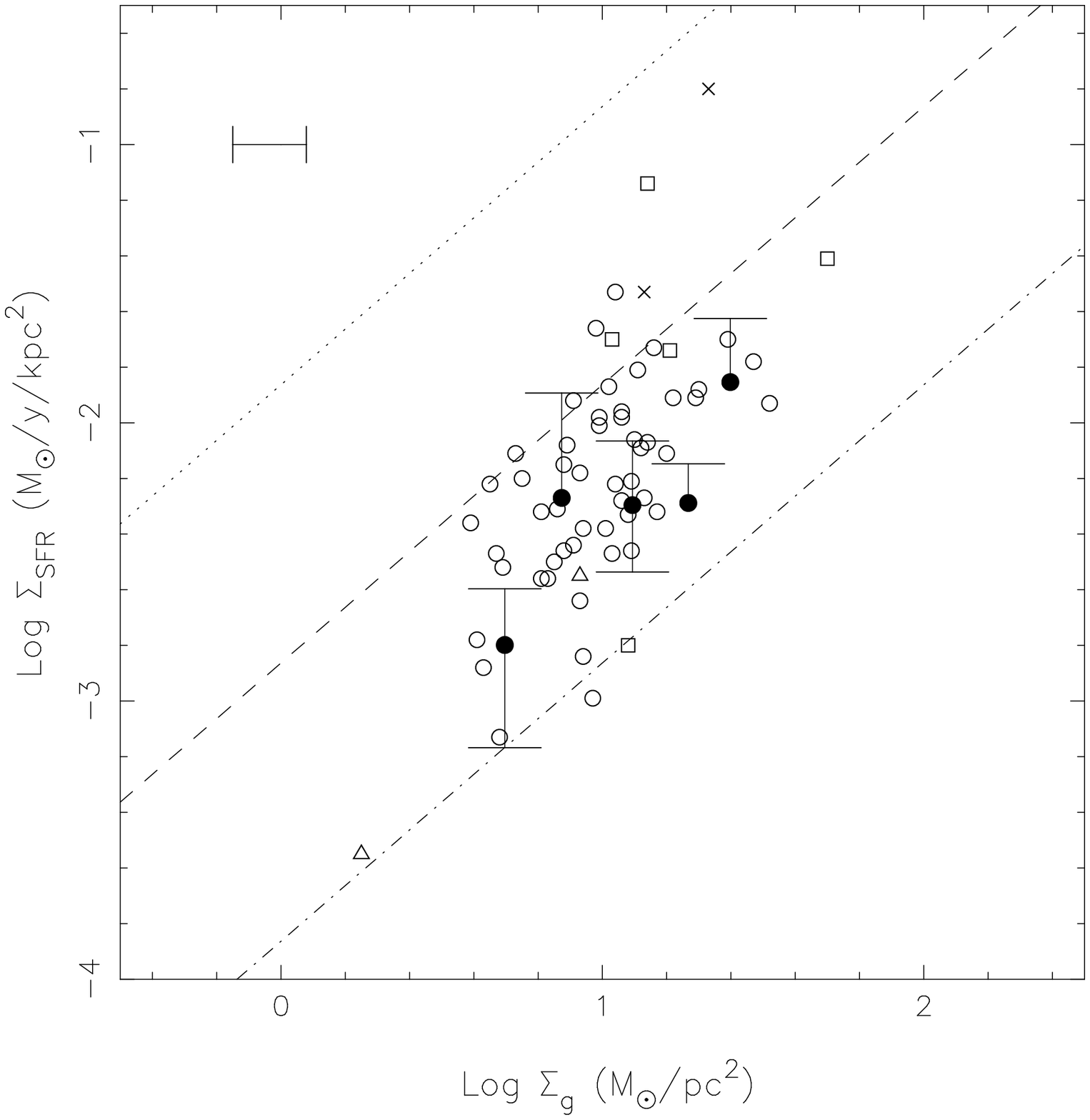}    
\caption{Disk-averaged SFR surface density ($\Sigma_{\rm SFR}$) as a function 
of average gas surface density ($\Sigma_{\rm g}$) for our galaxy sample and for
the sample of Kennicutt (1998) of 61 normal disk galaxies with SFR 
determined from H$_{\alpha}$ measurements. 
The 5 galaxies from our sample are plotted as filled circles 
and the corresponding
$\Sigma_{\rm SFR}$, $\Sigma_{\rm SFR}^{\rm min}$, 
$\Sigma_{\rm SFR}^{\rm max}$, $\Sigma_{\rm g}$ and disk areas are listed 
in Table~3 of Misiriotis et al. (2001). The $\Sigma_{\rm SFR}$ are derived from
the errors in the IRAS data, e.g. the upper and lower limits for the SFRs are 
calculated such that the predicted SED is still consistent with the IRAS 
colours (within the 20$\%$ IRAS error bars).
The SFR surface densities were calculated by averaging the SFRs  over disks 
with an optically defined boundary
($R_{\rm o}$) taken to be 3 times the intrinsic radial scalelength $h_{\rm s}$ 
determined from the radiation  transfer modelling in the I band. 
The galaxies from the sample of Kennicutt are plotted as open circles 
(Sb,Sc,SBb,SBc), triangles (Sa), open squares  (Unknown/Not Available),
crosses (Irr).
The dotted, dashed and dot-dashed lines represent
star-formation efficiencies corresponding to consumptions of 100,
10 and 1~percent of the gas mass in 10$^{8}\,$yr.}
\end{figure} 

Fig.~7 depicts the relation between the disk-averaged surface density in 
SFR ($\Sigma_{\rm SFR}$) as a 
function of the average gas surface density ($\Sigma_{\rm g}$) for the 
5 galaxies. The upper and lower limits for the SFRs are calculated such that 
the predicted SED is still consistent with the IRAS colours 
(within the 20$\%$ IRAS error bars). Lower error bars are not given when 
the plotted SFRs represent lower limits for the SFR (maximum values for the 
factors $F$, see discussion in Sect. 3 of Misiriotis et al. 2001). In these
cases lower limits would be
possible only if we allowed for different sources of uncertainties, like
variations in the spectral shape of the template used for the
HII regions. However this is hard to quantify, and in the following we assume
that the errors of the SFR are given only by the uncertainties in
the IRAS data. The plotted data are summarised in Table~3 of Misiriotis et
al. (2001) and details on the calculations of the error bars and of the gas
masses can be found in the same paper.
The horizontal error bar corresponds to the uncertainty in the gas masses 
for which we adopted an average 0.2 dex error. The surface area of the disk 
was calculated for $R_{\rm o}=(3 \pm 0.5) h_{\rm d}$, where $h_{\rm d}$ is 
the intrinsic radial scalelength determined from the radiation-transfer 
modelling in the I band.
In their analysis of surface photometry of
the outer regions of spiral disks, Pohlen et al. (2000) show that
the disk boundaries are typically in this range.

The points for the 5 galaxies in Fig.~10 lie within the
area of the diagram occupied by the galaxies in the Kennicutt sample. The 
match is even better for those members of the Kennicutt sample with  
Hubble types Sb to Sc.  
This agreement is quite reassuring, bearing in
mind the several factors which could introduce a systematic
difference between the SFRs inferred for a sample of nearly face-on systems
from H$_{\alpha}$ measurements compared with the present technique for
edge-on systems based on an analysis of broad-band non-ionising UV 
re-radiated in the FIR-submm range.

Firstly, the H$_{\alpha}$
analysis is sensitive to the most massive stars and in particular to the
assumed mass cut of 100\,M$_{\odot}$. Whereas the FIR-submm modelling also
assumes the same mass cut in the conversion of SFR to non-ionising luminosity,
our model is less sensitive to this effect.

Secondly, whereas the H$_{\alpha}$ is sensitive to the star-formation
history of the last 10$^{7}$yr, our broad-band FIR-submm SED analysis samples 
approximately the last 10$^{8}$yr. Thus, our analysis is consistent
with the basic hypothesis (see Kennicutt 1998) for ``normal'' spiral
galaxies of a steady star-formation activity. In principle, we could
extend our analysis based on our determinations of the intrinsic
populations and use the determined intrinsic colours
to determine more accurately the SFR history of the galaxies.

The assumption of a steady-state star-formation rate is also broadly
consistent with the timescales for the exhaustion of the current gas supply
under the derived SFRs.
The dotted, dashed and dot-dashed lines in Fig.~10 represent
star-formation efficiencies corresponding to consumptions of 100,
10 and 1~percent of the gas mass in 10$^{8}$\,yr.

Thirdly, the SFRs derived from H$_{\alpha}$ were corrected by a single
factor for extinction, despite the varying orientations. As well as 
possibly affecting the vertical position of the galaxies on the plot, 
this may induce some scatter, especially if all the dust were diffusely
distributed. The systematic effect may be expressed in terms of the factor $F$:
an overestimation of the factor $F$ is equivalent to an overestimation of the
local extinction in the star-formation regions (statistically averaged 
over the population of HII regions in a disk). Thus, while moving to higher  
factors $F$ would move the points for the 5 galaxies towards lower SFRs, it
would have the opposite effect for the SFRs determined from the H$_{\alpha}$.

Lastly, we remark that NGC~891 does not appear as an exceptional system
compared with the other 4 galaxies in our sample (and with Kennicutt's [1998]
normal galaxy sample) on the basis of SFR normalised to disk area. 
Our work thus provides no evidence that this galaxy's exceptional layer of 
extraplanar H$_{\alpha}$-emitting diffuse ionising gas 
(e.g. Hoopes et al. 1999) and surrounding 
X-ray-emitting hot gas (e.g. Bregman \& Houck 1997) is attributable to
unusual star-formation activity.

\section{Discussion and Outlook}

We have described a ``two-dust-disk'' model which can successfully account
for the observed optical-FIR/submm characteristics of
``normal'' edge-on spiral galaxies in terms of three fundamental
parameters - the $SFR$, $F$ - the fraction of non-ionising UV absorbed
locally in HII regions, and $M_{\rm dust}$ -  the mass of a second dust disk
associated with the young stellar population. We found these parameters to play 
 the key role in determining the amplitude and shape of the FIR-submm SED in 
galaxies. Some other
parameters were given larger attention in the literature, like for example
clumpiness, different grain populations, metallicity.

Clumpiness has been extensively discussed by Witt \& Gordon (1996, 2000) and
used by Gordon et al. (2001) and Matthews \& Wood (2001). 
However, the clumpiness in the interstellar medium is difficult to relate to 
observations. For example the optical properties of grains in clouds and 
fundamental parameters such as cloud size and density contrast are poorly 
known. Because of such uncertainties in
the parameterisation of clumpy media, our model does not explicitly
incorporate clumps, though our radiative transfer code can 
treat any arbitrary distribution of emitters and absorbers. However, our model 
considers
implicitly the clumpy distribution in form of the spectral template used for the
star-forming complexes. Thus, our model does not include clumps that are not
associated with stellar sources, the so called ``quiescent clouds'', but does
include clouds associated with star-forming regions. As
discussed by Popescu et al. (2000),  the quiescent clumps must be optically 
thick to the diffuse UV/optical radiation field in the disk to have an impact on
the predicted submm emission. They would then radiate at predominantly 
longer wavelengths than the diffuse disk emission in the FIR/submm spectral 
range. One can speculate that such optically thick ``quiescent clouds'' 
could be physically identified with partially or wholly collapsed clouds 
that, for lack of a trigger, have not (yet) begun to form stars. However,
due to the lack of intrinsic sources, a very substantial mass of dust would 
have to be associated with the quiescent clouds to account for a substantial
change in the submm SED of galaxies. On the other hand, there may be
also dark clouds associated with star-forming complexes.
In the Milky Way HII regions around newly born massive stars are commonly 
seen in juxtaposition to parent molecular clouds (e.g., M17). This is thought 
to be a consequence of the fragmentation of the clouds due to mechanical 
energy input from the winds of the massive stars. Thus, warm dust emission from
cloud surfaces directly illuminated by massive stars can be seen along
a fraction of the lines of sight, together with cold submm
dust emission from the interior of the associated optically thick cloud 
fragments. These dense clumps (with or without associated sources) may 
contribute to the submm emission, and thus supplement the contribution of
compact HII regions. Our template for localised sources was successful in
reproducing the FIR-submm surface brightness distribution for the prototypical 
galaxy NGC~891. However, modelling of more galaxies may be needed to check 
whether our
template is a good representation for localised sources, or whether it needs 
to be supplemented with a submm cold component of dark clouds. More detailed
FIR-submm observations of star-forming complexes may also help in improving
our HII-region templates.
 
Different dust populations can produce different types of extinction curves
and change the colour of the FIR SED. Our model used the Laor \& Draine (1993)
silicate/graphite mix, compatible with a Milky Way extinction law. Our choice
was based on the agreement between our model extinction curve and  the
empirical extinction law derived for the studied galaxies independently by 
Xilouris et al. (1997, 1998, 1999) via the radiative transfer modelling of the
optical images. In the case that different galaxies would require different 
extinction laws, we would adopt a different dust model 
(chemical composition, grain size distribution). 

Finally the metallicity was fixed in our model to be the solar
metallicity. Again, this choice was confirmed by the agreement in the
extinction curves. A different metallicity may also affect the redistribution
of the UV photons with UV wavelengths, as given by the stellar population 
synthesis models.
  
In the SED model presented here we did not consider any interactions or effects
of inflows/outflows in galaxies. Observationally there is however increased 
evidence for H$\alpha$ extraplanar emission in galaxies (e.g. 
Hoopes et al. 1999, Howk \& Savage 2000), of X-ray (e.g. 
Bregman \& Houck 1997) and radio halos (e.g. Allen et al. 1978), of 
galactic (starburst-driven) winds (e.g. Heckman et al. 2000) or for enhanced 
FIR emission in interacting/merging systems (e.g. Sanders et al. 1998). A 
discussion of the possible
contamination of the FIR disk emission from an extraplanar component has been
qualitatively discussed in Popescu et al. (2000b). Moreover, the effect of the 
cluster environment can also affect the morphology of both stellar and dust 
distribution, or can produce even more dramatic effects,  for example by 
sweeping the gas and dust material and producing tail-like companions with
collisionally heated grains visible in the FIR. As noted by Popescu et
al. (2000a), the FIR flux density of the transient IR
emission from the 
dust trail is predicted to rival that of the photon-heated dust in the 
galactic disk, and, despite the difference in heating mechanism,
have similar colours (with a spectral peak in the 100-200$\,{\mu}m$ range). 
This, combined with removal of dust from the disk, and hence reduced internal
extinction, will create a discrete system with brighter apparent blue 
magnitudes and a boosted spatially integrated IR flux density. If seen in a 
distant cluster, where the intracluster IR component could not be resolved from
the disk component, this could create the illusion of a galaxy with an 
enhanced star-formation activity, even though the star-formation in the
galaxy may actually be somewhat suppressed by the gas removal in reality.

Our model has been applied only to edge-on galaxies, where the 
scale heights of the old stellar population and old dust disk can be directly 
determined. However, our model will also be applicable to face-on systems 
where the scale heights cannot be so directly determined,
making use of UV data as an additional constraint.
Although our model requires resolved optical/NIR images to constrain
the old stellar population and associated dust, it relies on
geometry-sensitive colour information in the FIR/submm to constrain the
spatial distributions of young stars and associated dust.
The model will therefore be applicable to studies of cosmologically
distant ``normal'' galaxies, which, though detectable, will be unresolved
with forthcoming generations of spaceborne FIR observatories.
It is to be expected that the optical-FIR-submm SEDs of these objects
will differ systematically from their local universe counterparts,
not only due to the presence of younger stellar populations,
but also, for example, because of evolution of stellar disk thicknesses and 
changes in the dust abundance and composition.

SED modelling from UV to submm has an important application to interpretation
of the cosmological FIR background radiation in terms of constituent
galaxies. This particularly applies to the possible contribution from normal
galaxies in the early universe, since these are too faint to
be observed directly. The development of SED models incorporating a
self consistent physical theory for the evolution of these systems may
ultimately be needed for this application.

\section{Acknowledgements}

C. Popescu would like to express her gratitude to the organisers of the 
JENAM 2001 and AG meeting, and especially to Dr. Magdalena Stavinschi, for 
their invitation to give a Highlight Talk at the meeting. We are grateful to 
all collaborators in this project, N. Kylafis, A. Misiriotis, J. Fischera and 
B.F. Madore. We would also like to acknowledge E.M. Xilouris and J. Gallagher 
for interesting and constructive discussions.  

\vspace{0.7cm}
\noindent
{\large{\bf References}}
{\small

\bref 
Allen R.J., Baldwin J.E., Sancisi R., 1978, A\&A 62, 397

\bref
Alton P. B., Bianchi S., Rand R. J., et al., 1998, ApJ, 507, L125

\bref
Alton P. B., Rand R.J., Xilouris E. M., et al., 2000, A\&A, 356, 795

\bref
Bekki, K., \& Shioya, Y. 2000, ApJ, 542, 201

\bref
Bianchi, S.,  Davies, J. I., \& Alton, P. B. 2000, A\&A, 359, 65


\bref 
Bregman J.N., \& Houck, J. C. 1997, ApJ, 485, 159

\bref
Bruzual, A. G., \& Charlot, S.  2001, in preparation

\bref
Dale, D. A., Helou, G., Contursi, A., Silbermann, N. A. \& Kolhatkar, S. 2001,
ApJ, 549, 215

\bref
Devriendt, J. E. G., Guiderdoni, B. \& Sadat, R. 1999, A\&A, 350, 381

\bref
D\'esert, F. X., Boulanger, F., \& Puget, J. L. 1990, A\&A, 237, 215

\bref 
Draine B. T. \& Lee H. M. 1984, ApJ, 285, 89

\bref 
Draine B. T., \&  Anderson N. 1985, ApJ, 292, 494


\bref 
Dwek E. 1986, ApJ, 302, 363

\bref
Efstathiou, A., Rowan-Robinson, M., \& Siebenmorgen, R., 2000, MNRAS, 313, 375

\bref
Ferrara, A., Bianchi, S., Cimatti, A., \& Giovanardi, C. 1999, ApJS, 123, 437




\bref
Gordon, K. D., Misselt, K. A., Witt, A. N., \& Clayton, G. C. 2001, ApJ, 551, 
269

\bref
Guhathakurta P., \& Draine B. T. 1989, ApJ, 345, 230

\bref
Gu\'elin, M., Zylka, R., Mezger, P. G., et al., 1993, A\&A, 279, L37

\bref
Heckman, T. M., Lehnert, M. D., Strickland, D. K., \& Armus, L. 2000, ApJS, 
129, 493 

\bref  
Hoopes C. G., Walterbos R. A. M., \& Rand R. J., 1999, ApJ 522, 669

\bref 
Howk J. C., \& Savage, B. D., 2000, AJ, 119, 644


\bref
Kennicutt, R. C. Jr. 1998, ApJ, 498, 541

\bref 
Kuchinski, L. E., Terndrup, D. M., Gordon, K. D., Witt, A. N. 
1998, AJ, 115, 1438  

\bref
 Kylafis, N. D., \& Bahcall, J. N. 1987, ApJ, 317, 637

\bref
Laor A., \& Draine B. T. 1993, ApJ 402, 441

\bref
Mathis, J. S., Rumple, W., \& Nordsieck, K. H. 1977, ApJ, 217, 425

\bref 
Matthews, L. D., \& Wood, K. 2001, ApJ 548, 150

\bref
Misiriotis A., Kylafis N. D., Papamastorakis J. \&  Xilouris E. M.
   2000, A\&A, 353, 117

\bref
Misiriotis A., Popescu C. C., Tuffs R.J . \& Kylafis, D. 2001, A\&A, 372, 775

\bref
Ohta, K. \& Kodaira, K. 1995, PASJ, 47, 17

\bref
Pohlen, M., Dettmar, R.-J., Lutticke, R. 2000, A\&A, 357, L1

\bref
Popescu, C. C., Tuffs, R. J., Fischera, J. \& V\"olk, H. J. 2000a, A\&A, 
354, 480

\bref
Popescu, C. C., Misiriotis, A., Kylafis, N. D., Tuffs, R. J. \& Fischera, J. 
2000b, A\&A, 362, 138

\bref
Popescu, C. C.,  Madore, B. F., Tuffs, R. J., \& Kylafis, N. D. 2001, 
AAS198, 76.01

\bref
Popescu, C. C., Tuffs, R. J., V\"olk, H. J., Pierini, D., 
\& Madore, B. F.  2002, ApJ, 567, 221


\bref
Sanders, D. B., Soifer, B. T., Elias, J. H., et al., 1998, ApJ, 325, 74
 
\bref
Silva, L., Granato, G .L., Bressan, A. \& Danese, L., 1998, ApJ, 509, 103

\bref
Swaters, R.A., Sancisi, R. \& van der Hulst, J.M. 1997, ApJ, 491, 140

\bref
Tuffs, R. J., Popescu, C. C., Pierini, D., V\"olk, H. J., Hippelein,
H., et al., 2002, ApJS, 139, 37

\bref
Weingartner, J. C., \& Draine, B. T. 2001, ApJ, 548, 296

\bref
Witt, A. N., \& Gordon, K. G. 1996, ApJ, 463, 681

\bref
Witt, A. N., \& Gordon, K. G. 2000, ApJ, 528, 799

\bref 
Xilouris, E. M., Kylafis, N. D., Papamastorakis, J., Paleologou \& 
E. V., Haerendel, G., 1997, A\&A, 325, 135

\bref 
Xilouris, E. M., Alton, P. B., Davies, J. I., et al., 1998, A\&A, 331, 894

\bref
Xilouris, E. M., Byun, Y. I., Kylafis, N. D., 
Paleologou, E. V., Papamastorakis, J., 1999, A\&A, 344, 868

\bref
Xu, C. \& Buat, V., 1995, A\&A, 293, L65

\bref Xu, C. \& Helou, G., 1996, ApJ, 456, 163

\bref
Young, J. S., Xie, S., Kenney, J. D. P. \& Rice, W. L., 1989, ApJS, 70, 699
}

\vfill

\end{document}